\newif\ifFull
\Fulltrue
\ifFull
\documentclass[11pt]{article}
\usepackage{fullpage}
\else
\documentclass[10pt]{llncs}
\fi
\usepackage{times}
\usepackage{listings}
\ifFull
\newtheorem{theorem}{Theorem}
\newtheorem{corollary}{Corollary}
\newtheorem{lemma}{Lemma}

\newenvironment{proof}{\noindent{\bf Proof:}}{\hspace*{\fill}\rule{6pt}{6pt}\bigskip}
\fi
\newcommand{\mod}{{\rm mod\ }}
\makeatletter
\def\@begintheorem#1#2{\sl \trivlist \item[\hskip \labelsep{\bf #1\ #2:}]}
\def\@opargbegintheorem#1#2#3{\sl \trivlist
      \item[\hskip \labelsep{\bf #1\ #2\ #3:}]}
      \makeatother

\begin{document}
\ifFull
\title{\LARGE Improved Combinatorial Group Testing\\ 
           \LARGE Algorithms for Real-World Problem Sizes}
\date{}
\else
\title{\LARGE Improved Combinatorial Group Testing\\ 
           \LARGE for Real-World Problem Sizes}
\fi
\author{
{David Eppstein}
\and
{Michael T. Goodrich}
\and
{Daniel S. Hirschberg}
}
\ifFull
\else
\institute{Dept.~of Computer Science, 
    Univ.~of California, Irvine, CA 92697-3425 USA.
    \email{\{eppstein,goodrich,dan\}(at)ics.uci.edu.}}
\fi
\maketitle

\ifFull
\begin{center}
Dept.~of Computer Science \\
    Univ.~of California, Irvine \\
    Irvine, CA 92697-3425 USA \\
    \texttt{\{eppstein,goodrich,dan\}(at)ics.uci.edu.}
\end{center}

\bigskip
\fi 

\begin{abstract}
We study practically efficient methods for performing combinatorial
group testing.  We present efficient non-adaptive and two-stage
combinatorial group testing algorithms, which identify the at most $d$
items out of a given set of $n$ items that are defective, using fewer
tests for all practical set sizes.  For example, our two-stage
algorithm matches the information theoretic lower bound for the number
of tests in a combinatorial group testing regimen.

\end{abstract}

\section{Introduction}
The problem of combinatorial group testing dates back to World War~II, 
for the problem of determining which in a group of $n$ blood samples
contain the syphilis antigen (hence, are contaminated).
Formally, in combinatorial group testing, we are given a set of $n$ items,
at most $d$ of which are defective (or contaminated),
and we are interested in identifying exactly which of the $n$ items
are defective.
In addition, items can be ``sampled'' and these samples can be ``mixed''
together, so tests for contamination can be applied to arbitrary subsets
of these items.
The result of a test may be positive, indicating that
at least one of the items of that subset is defective,
or negative, indicating that all items in that subset are good.
Example applications that fit this framework include:
\begin{itemize}
\item \emph{Screening blood samples for diseases.} 
In this application,
items are blood samples and tests are disease
detections done on mixtures taken from selected samples.
\item \emph{Screening vaccines for contamination.} 
In this case, items are vaccines and tests are
cultures done on mixtures of samples taken from selected
vaccines.
\item \emph{Clone libraries for a DNA sequence.}
Here, the items are DNA subsequences (called
\emph{clones}) and tests are done on pools of clones to determine
which clones contain a particular DNA sequence (called a
\emph{probe})~\cite{fkkm-gtpse-97}.
\item \emph{Data forensics.}
In this case,
items are documents and the tests are
applications of one-way hash functions with known expected values
applied to selected collections of documents.
\ifFull
The differences from
the expected values are then used to identify which, if any, of the documents
have been altered.
\fi
\end{itemize}

The primary goal of a testing algorithm is to identify all defective items
using as few tests as possible. 
That is, we wish to minimize the following function:
\begin{itemize}
\item $t(n,d)$: The number of tests needed to
identify up to $d$ defectives among $n$ items.
\end{itemize}
This minimization may be subject to possibly additional constraints, as well.
For example, we may wish to identify all the defective items in a
single (\emph{non-adaptive})
round of testing, we may wish to do this in two
(\emph{partially-adaptive}) rounds, 
or we may wish to perform the tests sequentially one after the other
in a \emph{fully adaptive} fashion.

In this paper we are interested in efficient solutions to
combinatorial group testing problems for realistic problem sizes, 
which could be applied to solve the motivating examples given above.
That is, we wish solutions that minimize $t(n,d)$ for practical
values of $n$ and $d$ as well as asymptotically.
Because of the inherent delays that are built into fully adaptive,
sequential solutions, we are interested only in solutions
that can be completed in one or two rounds.
Moreover, 
we desire solutions that are efficient not only in terms of
the total number of tests performed, but also for the following
measures:
\begin{itemize}
\item $A(n,t)$: The \emph{analysis} time needed to determine which
	items are defective.
\item $S(n,d)$: The \emph{sampling} rate---the maximum number
of tests any item may be included in.
\end{itemize}
An analysis algorithm is said to be {\it efficient} if $A(n,t)$ is
$O(tn)$, where $n$ is the number of items
and $t$ is the number of tests conducted.
It is {\it time-optimal} if $A(n,t)$ is $O(t)$.
Likewise, 
we desire efficient sampling rates for our algorithms; that is,
we desire that $S(n,d)$ be $O(t(n,d)/d)$.
Moreover, we are interested in this paper in solutions that improve
previous results, either asymptotically or by constant factors, 
for realistic problem sizes.
We do not define such ``realistic'' problem sizes formally, but we
may wish to consider as unrealistic a problem that is larger than
the total memory capacity (in bytes) of all CDs and DVDs in the world
($<10^{25}$), the number of atomic particles in the earth
($<10^{50}$), or the number of atomic particles in the universe
($<10^{80}$).

\paragraph{Viewing Testing Regimens as Matrices.}
%
%
%
%
%
%
A single round in a combinatorial group testing algorithm 
consists of a test regimen and an analysis algorithm
(which, in a non-adaptive (one-stage) algorithm, 
must identify all the defectives).
The test regimen can be modeled by a $t \times n$ Boolean matrix, $M$.
Each of the $n$ columns of $M$ corresponds to one of the $n$ items.
Each of the $t$ rows of $M$ represents a test of items whose
corresponding column has a 1-entry
in that row.
All tests are conducted before the results of any test is made available.
The analysis algorithm uses the results of the $t$ tests
to determine which of the $n$ items are defective.

As described by Du and Hwang \cite{dh-cgtia-00}(p.~133),
the matrix $M$ is $d$-{\it disjunct} if the Boolean sum of
any $d$ columns does not contain any other column.
In the analysis of a $d$-{\it disjunct} testing algorithm,
items included in a test with negative outcome
can be identified as pure.  Using a $d$-disjunct matrix
enables the conclusion that if there are $d$ or fewer items
that cannot be identified as pure in this manner then all those items
must be defective and there are no other defective items.
If more than $d$ items remain then at
least $d+1$ of them are defective.
Thus, using a $d$-disjunct matrix enables an
efficient analysis algorithm, with $A(n,t)$ being $O(tn)$.

$M$ is $d$-{\it separable} ($\overline d$-{\it separable})
if the Boolean sums of $d$ (up to $d$) columns are all distinct.
The $\overline d$-separable property implies that
each selection of up to $d$ defective items induces a different set
of tests with positive outcomes.  Thus, it is possible to identify
which are the up to $d$ defective items by checking, for each
possible selection, whether its induced positive test set is exactly the
obtained positive outcomes.  However, it might not be possible
to detect that there are more than $d$ defective items.
This analysis algorithm takes time $\Theta(n^d)$
or requires a large table mapping $t$-subsets to $d$-subsets.

Generally, $\overline d$-separable matrices can be
constructed with fewer rows than can $d$-disjunct matrices
having the same number of columns.
Although the analysis algorithm described above for
$d$-separable matrices is not efficient,
some $\overline d$-separable matrices that are not
$d$-disjunct have an efficient analysis algorithm.

\paragraph{Previous Related Work.}
Combinatorial group testing is a rich research area with many
applications to many other areas, including
communications, cryptography, and networking~\cite{colbourn99applications}.
For an excellent discussion of this topic,
the reader is referred to the book by 
Du and Hwang~\cite{dh-cgtia-00}.
For general $d$,
Du and Hwang \cite{dh-cgtia-00}(p.~149) describe a slight modification
of the analysis of a construction due to Hwang and S\'{o}s \cite{HS87}
that results in a $t\times n$ 
$d$-disjunct matrix, with $n \geq (2/3)3^{t/16d^2}$,
and so $t \leq 16d^2(1 + \log_{3}2 + (\log_{3}2) \lg n)$.
For two-stage testing, Debonis \textit{et al.}~\cite{dgv-gfsao-03}
provide a scheme that achieves a number of tests within 
a factor of $7.54(1+o(1))$ of the information-theoretic lower 
bound of $d\log (n/d)$.
For $d=2$,
Kautz and Singleton \cite{KS64} construct
a 2-disjunct matrix with $t = 3^{q+1}$ and
$n = 3^{2^q}$, for any positive integer $q$.
Macula and Reuter \cite{MR98} describe a $\overline 2$-separable
matrix and a time-optimal analysis algorithm with
$t = (q^2 +3q)/2$ and $n = 2^q - 1$, for any positive integer $q$.
For $d=3$, Du and Hwang \cite{dh-cgtia-00}(p.~159)
describe the construction of a $\overline 3$-separable matrix
(but do not describe the analysis algorithm) with
$t = 4 {3q \choose 2} = 18q^2 - 6q$ and $n = 2^q-1$,
for any positive integer $q$.

\paragraph{Our Results.}
In this paper, we consider problems of identifying defectives using
non-adaptive or two-stage protocols with efficient analysis algorithms.
We present several such algorithms that require fewer tests
than do previous algorithms for practical-sized sets, although we omit
the proofs of some supporting lemmas in this paper, 
due to space constraints.
Our general case algorithm, which is based on a method we call the
Chinese Remainder Sieve, improves
the construction of Hwang and S\'{o}s \cite{HS87}
for all values of $d$ for real-world problem instances as well as for
$d\ge n^{1/5}$ and $n\ge e^{10}$.
Our two-stage algorithm
achieves a bound for $t(n,d)$ that is within a factor of $4(1+o(1))$ 
of the information-theoretic lower bound. This bound improves the
bound achieved by Debonis \textit{et al.}~\cite{dgv-gfsao-03} by
almost a factor of $2$.
Likewise, our algorithm for $d=2$ improves on the number of tests
required for all real-world problem sizes and is time-optimal
(that is, with $A(n,t)\in O(t)$).
Our algorithm for $d=3$ is the first known
time-optimal testing algorithm for that $d$-value.
Moreover, our algorithms all have efficient sampling rates.

\section{The Chinese Remainder Sieve}
In this section, we present a solution to the 
problem for determining which items are defective
when we know that there are at most $d<n$ defectives.
Using a simple number-theoretic method,
which we call the \emph{Chinese Remainder Sieve} method,
we describe the construction of a $d$-disjunct matrix
with $t = O(d^2 \log^2 n / (\log d + \log \log n))$.
As we will show,
our bound is superior to that of the method of Hwang and S\'{o}s \cite{HS87},
for all realistic instances of the combinatorial group testing
problem.

Suppose we are given $n$ items, numbered $0,1,\ldots,n-1$, such that
at most $d<n$ are defective.
Let \{$p_1^{e_1},p_2^{e_2},\ldots,p_k^{e_k}$\} be a 
sequence of powers of distinct primes,
multiplying to at least $n^d$.
That is, $\prod_{j} p_j^{e_j} \ge n^d$.
We construct a $t \times n$ matrix $M$ as the vertical concatenation of 
$k$ submatrices, $M_1,M_2,\ldots,M_k$.
Each submatrix $M_j$ is a $t_j\times n$ testing matrix, where
$t_j=p_j^{e_j}$; hence, $t=\sum_{j=1}^k p_j^{e_j}$.
We form each row of $M_j$ by associating it with a non-negative value $x$ less
than $p_j^{e_j}$.
Specifically,
for each $x$, $0\le x< p_j^{e_j}$, form a test in $M_j$ consisting of the
item indices (in the range $0,1, \ldots, n-1$) that equal 
$x\, (\mod p_j^{e_j})$.
For example, if $x=2$ and $p_j^{e_j}=3^2$, then the row for $x$ in
$M_j$ has a $1$ only in columns $2$, $11$, $20$, and so on.

The following lemma shows that the test matrix $M$ is $d$-disjunct.

\begin{lemma}
\label{lem:6}
If there are at most $d$ defective items, and all tests in $M$ 
are positive for $i$, then $i$ is defective.
\end{lemma}

\begin{proof}
If all $k$ tests for $i$ 
(one for each prime power $p_j^{e_j}$)
are positive, then there exists at least one defective item.
With each positive test that includes $i$ (that is, it has a $1$ in
column $i$), let $p_j^{e_j}$ be the modulus used for this test, 
and associate with $j$ a defective index $i_j$ that was
included in that test (choosing $i_j$ arbitrarily in case test $j$ includes
multiple defective indices).  
\ifFull
For any defective index $i'$, let
$$P_{i'}=\prod_{j\mbox{ s.t. }i_j=i'} p_j^{e_j}.$$
\else
For any defective index $i'$, let
$P_{i'}=\prod_{j\mbox{ s.t. }i_j=i'} p_j^{e_j}$.
\fi
That is, $P_{i'}$ is the product of all the prime powers such
that $i'$ caused a positive test that included $i$ for that prime
power.
Since there are $k$
tests that are positive for $i$,
each $p_j^{e_j}$ appears in exactly one of these products, $P_{i'}$.
So $\prod P_{i'}=\prod p_j^{e_j}\ge n^d$.
Moreover, there are at most $d$ products, $P_{i'}$.
Therefore, $\max_{i'} P_{i'}\ge (n^d)^{1/d}=n$; hence,
there exists at least one defective index ${i'}$
for which $P_{i'}\ge n$.
By construction, $i'$ is congruent to the same values
to which $i$ is congruent, modulo each of the prime
powers in $P_{i'}$.  
By the Chinese Remainder Theorem, the solution to
these common congruences is unique modulo the least common multiple
of these prime powers, which is $P_{i'}$ itself.
Therefore, $i$ is equal to ${i'}$ modulo a
number that is at least $n$, so $i={i'}$; hence, $i$ is defective.
\end{proof}

The important role of the Chinese Remainder Theorem in the proof of
the above lemma gives rise to our name for this construction---the
Chinese Remainder Sieve.

\paragraph{Analysis.}
As mentioned above,
the total number of tests, $t(n,d)$,
constructed in the Chinese Remainder Sieve is $\sum_{j=1}^k p_{j}^{e_j}$,
where $\prod p_j^{e_j} \ge n^d$.
If we let each $e_j=1$, 
we can simplify our analysis to note that
$t(n,d)=\sum_{j=1}^k p_{j}$, where $p_j$ denotes the $j$-th prime number and
$k$ is chosen so that $\prod_{j=1}^k p_j \ge n^d$.
To produce a closed-form upper bound for $t(n,d)$, we make use of the
prime counting function,
$\pi(x)$, which is the number of primes less than or equal to $x$.
We also use the well-known \emph{Chebyshev function},
$
\theta(x)=\sum_{j=1}^{\pi(x)}\ln p_j 
$.
In addition, we make use of the following (less well-known)
prime summation function,
$
\sigma(x) = \sum_{j=1}^{\pi(x)} p_j 
$.
Using these functions, we bound the number of tests in the Chinese
Remainder Sieve method as
$t(n,d) \le \sigma (x)$, where $x$ is chosen so that 
$\theta(x)\ge d\ln n$, since
$\ln \prod_{p_j\le x} p_j = \theta(x)$.
For the Chebyshev function, it can be shown~\cite{bs-ant-96} that
$\theta(x)\ge x/2$ for $x>4$ and that $\theta(x) \sim x$ for large $x$.
So if we let $x = \lceil 2d\ln n\rceil$, then $\theta(x)\ge d\ln n$.
Thus, we can bound the number of tests in our method as
$t(n,d) \le \sigma (\lceil 2d\ln n\rceil)$.
To further bound $t(n,d)$, we use the following lemma, which may be
of mild independent interest.

\begin{lemma}
\label{lemma-sigma}
For integer $x\ge 2$,
\[
\sigma(x) < \frac{x^2}{2\ln x} \left(1 + \frac{1.2762}{\ln x}\right) .
\]
\end{lemma}
\begin{proof}
Let $n=\pi(x)$.
Dusart~\cite{d-eefc-98,d-kpgtk-99} shows that, for $n\ge 799$,
\ifFull
\[
\frac{1}{n} \sum_{j=1}^n p_j < \frac{1}{2} p_n,
\]
\else
$({1}/{n}) \sum_{j=1}^n p_j < p_n/2$;
\fi
that is, the average of the first $n$ primes is half the value of the
$n$th prime.
Thus,
\[
\sigma(x) = \sum_{j=1}^{\pi(x)} p_j < \frac{\pi(x)}{2} p_n
\le \frac{\pi(x)}{2} x ,
\]
for integer $x\ge 6131$ (the $799$th prime).
Dusart~\cite{d-eefc-98,d-kpgtk-99} also shows that
\[
\pi(x) < \frac{x}{\ln x}\left(1+\frac{1.2762}{\ln x}\right) ,
\]
for $x\ge 2$.
Therefore, for integer $x\ge 6131$,
\[
\sigma(x) < \frac{x^2}{\ln x}\left(1+\frac{1.2762}{\ln x}\right) .
\]
In addition,
we have verified by an exhaustive computer search that this inequality
also holds for all integers $2\le x< 6131$.
This completes the proof.
\end{proof}

Thus, we can characterize the Chinese Remainder Sieve method as
follows.

\begin{theorem}
Given a set of $n$ items, at most $d$ of which are defective, the
Chinese Remainder Sieve method can identify the defective items using
a number of tests 
\[
t(n,d) < \frac{\lceil 2d\ln n\rceil^2}{2\ln \lceil 2d\ln n\rceil} 
         \left(1 + \frac{1.2762}{\ln \lceil 2d\ln n\rceil}\right) .
\]
\ifFull
The sample rate can be bounded by
\[
S(n,d) < \frac{\lceil 2d\ln n\rceil}{2\ln \lceil 2d\ln n\rceil} 
         \left(1 + \frac{1.2762}{\ln \lceil 2d\ln n\rceil}\right) ,
\]
and the analysis time, $A(n,t)$, is $O(n t(n,d))$.
\fi
\end{theorem}

By calculating the exact numbers of tests required
by the Chinese Remainder Sieve method for particular parameter values
and comparing these numbers to the claimed bounds for 
Hwang and S\'{o}s~\cite{HS87},
we see that our algorithm is an improvement when:
\begin{center}
\begin{tabular}{cc}
$\bullet$ $d=2$ and $n\le 10^{57}$ &
$\bullet$ $d=3$ and $n\le 10^{66}$ \\
$\bullet$ $d=4$ and $n\le 10^{70}$ &
$\bullet$ $d=5$ and $n\le 10^{74}$ \\
$\bullet$ $d=6$ and $n\le 10^{77}$ &
$\bullet$ $d\ge 7$ and $n\le 10^{80}$.
\end{tabular}
\end{center}

Of course, these are the most likely cases for any expected
actual instance of the combinatorial group testing problem.
In addition, our analysis shows that our method is superior to the
claimed bounds of Hwang and S\'{o}s~\cite{HS87} for
$d\ge n^{1/5}$ and $n\ge e^{10}$.
Less precisely, we can say that
$t(n,d)$ is $O(d^2 \log^2 n / (\log d + \log \log n))$,
that $S(n,d)$ is $O(d\log n/(\log d + \log\log n)$,
and
$A(n,t)$ is $O(tn)$, which is $O(d^2n \log^2 n / (\log d + \log \log n))$.

\ifFull
\paragraph{Heuristic Improvements.}
Although it will not reduce the asymptotic complexity of $t$,
we can reduce the number of tests by 
starting with a sequence of primes up to some upper bound $x$, and
efficiently constructing a set of good prime powers from this sequence.
We can allow some powers, $e_j$, to be zero (meaning that we don't
use this prime),
while giving others values greater than one. 
The objective is to choose carefully the values
$e_j$ in order to minimize the number of tests while maintaining the
property that $\prod p_j^{e_j}\ge n^d$.
This typically yields a savings of between five and ten percent.

An example implementation in Python 2.3 is shown
in the Appendix
in Figures~\ref{fig:Primes-subs}~and~\ref{fig:Primes}.
This implementation starts with the $e_j=1$ solution
to determine an initial suitable sequence of primes, $p_j$, to use.
It then does a backtracking search to find the optimal set of $e_j$
for these $p_j$, subject to the constraint that each $p_{j}^{e_j}$
is not greater than the largest prime in the original solution
(with each $e_j=1$).
Since the number of $e_j$ powers is sublogarithmic,
and most of them must be 0 or 1,
this backtracking search takes time sublinear in $n$ for fixed $d$.

\paragraph{Comparison of the Number of Tests Required.}
Table~\ref{tbl:Compared} lists the number of tests required by
the Hwang/S\'{o}s algorithm, our general algorithm
(using the initial set of primes $p_j$ having exponents $e_j=1$),
and our improved backtrack algorithm,
for some values of $n$.
As can be seen, for moderate values of $n$
our algorithms require a small fraction
of the number of tests required by the HS algorithm.
However, asymptotically for fixed $d$,
the HS algorithm requires fewer tests.

\begin{table}[htp]
\caption[]{Comparing $t(n)$ for $d=5$ and $d=10$}
\[
\begin{array}{|l|rrrrrrr|}
\hline
(d=5)           & 100 & 10^4 & 10^6 & 10^8 & 10^{10}& 10^{20}& 10^{30}\\
\hline
{\rm our\ bktrk}& 131 &  378 &  738 & 1176 &   1709 &   5737 & 11782 \\
{\rm our\ genl} & 160 &  440 &  791 & 1264 &   1851 &   6081 & 12339 \\
{\rm HS}        &2329 & 4006 & 5683 & 7359 &   9036 &  17420 & 25803 \\
\hline
\hline
(d=10)          & 100 & 10^4 & 10^6 & 10^8 & 10^{10}& 10^{20}& 10^{30}\\
\hline
{\rm our\ bktrk}& 378 & 1176 & 2350 & 3896 &   5737 &  19681 & 41020 \\
{\rm our\ genl} & 440 & 1264 & 2584 & 4227 &   6081 &  20546 & 42468 \\
{\rm HS}        &9316 &16023 &22730 &29437 &  36144 &  69678 &103213 \\
\hline
\end{array}
\]
\label{tbl:Compared}
\end{table}
\fi

\section{A Two-Stage Rake-and-Winnow Protocol}
In this section, we present a randomized construction for two-stage
group testing. 
This two-stage method uses
a number of tests within a constant factor of the information-theoretic
lower bound.  It improves previous 
upper bounds~\cite{dgv-gfsao-03}
by almost a factor of $2$.
In addition, it has an efficient sampling rate, with $S(n,d)$ being
only $O(\log (n/d))$.
All the constant factors ``hiding'' behind the big-ohs 
in these bounds are small.

\paragraph{Preliminaries.}
One of the important tools we use in our analysis is the following
lemma for bounding the tail of a certain distribution.
It is a form of Chernoff bound~\cite{mr-ra-95}.

\begin{lemma}
\label{lem:chernoff}
Let $X$ be the sum of $n$ independent indicator random
variables, such that $X=\sum_{i=1}^n X_i$,
where each $X_i=1$ with probability $p_i$, for
$i=1,2,\ldots,n$.
If $E[X]=\sum_{i=1}^n p_i \le{\hat\mu}<1$,
then, for any integer $k>0$,
\ifFull
\[
\Pr(X\ge k) \le \left(\frac{e{\hat\mu}}{k}\right)^k .
\]
\else
$\Pr(X\ge k) \le \left({e{\hat\mu}}/{k}\right)^k$.
\fi
\end{lemma}
\begin{proof}
Let $\mu=E[X]$ be the actual expected value of $X$.
Then, by a well-known Chernoff bound~\cite{mr-ra-95},
for any $\delta>0$,
\[
\Pr[X\ge(1+\delta)\mu] \le
  \left[\frac{e^{\delta}}{(1+\delta)^{1+\delta}}\right]^{\mu} .
\]
(The bound in \cite{mr-ra-95} is for strict inequality, but the same
bound holds for nonstrict inequality.)
We are interested
in the case when $(1+\delta)\mu=k$,
that is, when $1+\delta=k/\mu$.
Observing that $\delta < 1+\delta$, we can therefore deduce that
\[
\Pr(X\ge k) \le
  \left[\frac{e^{k/\mu}}{(k/\mu)^{k/\mu}}\right]^{\mu} =
\frac{e^k}{(k/\mu)^k} =
\left(\frac{e{\mu}}{k}\right)^k .
\]
Finally, noting that $\mu\le {\hat\mu}$,
\ifFull
\[
\Pr(X\ge k) \le \left(\frac{e{\hat\mu}}{k}\right)^k .
\]
\else
$\Pr(X\ge k) \le \left({e{\hat\mu}}/{k}\right)^k$.
\fi
\end{proof}

\ifFull
In addition to this lemma, we also use the following.
\fi

\begin{lemma}
\label{lem:choose}
If $d<n$, then
\ifFull
\[
{n \choose d} < \left(\frac{en}{d}\right)^d .
\]
\else
${n \choose d} < \left({en}/{d}\right)^d $.
\fi
\end{lemma}
\ifFull
\begin{proof}
\begin{eqnarray*}
{n \choose d} &=& \frac{n!}{(n-d)!\,d!} \\
              &=& \frac{n(n-1)(n-2)\cdots (n-d+1)}{d!} \\
	      &<& \frac{n^d}{d!} .
\end{eqnarray*}
By Stirling's approximation~\cite{clr-ia-90},
\[
d! = \sqrt{2\pi n} \left(\frac{d}{e}\right)^d (1+\theta(1/n)) .
\]
Thus, $d!> (d/e)^d$.
Therefore,
\[
\frac{n^d}{d!} <  \frac{n^d}{(d/e)^d} = \left(\frac{en}{d}\right)^d .
\]
\end{proof}
\fi

\paragraph{Identifying Defective Items in Two Stages.}
As with our Chinese Remainder Sieve method,
our randomized combinatorial group testing construction is based on the
use of a Boolean matrix $M$ where columns correspond to items and rows
correspond to tests, so that if $M[i,j]=1$, then item $j$ is included
in test $j$.
Let $C$ denote the set of columns of $M$.
Given a set $D$ of $d$ columns in $M$, and a specific column $j\in C-D$,
we say that $j$ is \emph{distinguishable} from $D$ if there is a row
$i$ of $M$ such that $M[i,j]=1$ but $i$ contains a $0$ in each of the
columns in $D$.
Such a property is useful in the context of group testing,
for the set $D$ could correspond to the defective items and
if a column $j$ is
distinguishable from the set $D$, then there would be a test
in our regimen that would determine that the item corresponding to
column $j$ is not defective.

An alternate and equivalent definition~\cite{dh-cgtia-00}(p.~165) 
for a matrix $M$ to be \emph{$d$-disjunct}
is if, for any $d$-sized subset $D$ of $C$, each
column in $C-D$ is distinguishable from $D$.
Such a matrix determines a powerful group testing regimen, but,
unfortunately, building such a matrix
requires $M$ to have 
$\Omega(d^2\log n/\log d)$ rows,
by a result of
Ruszink{\'o}~\cite{R94} (see also \cite{dh-cgtia-00}, p.~139).
The best known constructions
have $\Theta(d^2\log (n/d))$ rows \cite{dh-cgtia-00},
which is a factor of $d$ greater than information-theoretic lower
bound, which is $\Omega(d\log (n/d))$.

Instead of trying to use a matrix $M$ to determine all the defectives
immediately, we will settle for a weaker property for $M$, which
nevertheless is still powerful enough to define a good group testing
regimen.
We say that $M$ is \emph{$(d,k)$-resolvable} if, for any $d$-sized subset
$D$ of $C$, there are fewer than $k$ columns in $C-D$ that are not
distinguishable from $D$.
Such a matrix defines a powerful group testing regimen,
for defining tests
according to the rows of a $d$-resolvable matrix allows us to restrict
the set of defective items to a group $D'$ of smaller than $d+k$ size.  
Given this set, we can then perform an additional round of
individual tests on all the items in $D'$.
This two-stage approach is sometimes called the trivial two-stage
algorithm; we refer to this
two-stage algorithm as the \emph{rake-and-winnow} approach.

Thus, a $(d,k)$-resolvable matrix determines a powerful group testing
regimen.
Of course, a matrix is $d$-disjunct if and only if it is $(d,1)$-resolvable.
Unfortunately, as mentioned above, constructing a $(d,1)$-resolvable
matrix requires
that the number of rows (which correspond to tests) be significantly
greater than the information theoretical lower bound.
Nevertheless, if we are willing to use a $(d,k)$-resolvable matrix,
for a reasonably small value of $k$, we can come within a constant
factor of the information theoretical lower bound.

Our construction of a $(d,k)$-resolvable matrix is based on a simple,
randomized \emph{sample-injection} strategy,
which itself is based on the approach
popularized by the Bloom filter~\cite{B70}.
This novel approach also allows us to provide a strong worst-case
bound for the sample rate, $S(n,d)$, of our method.
Given a parameter $t$, which is a multiple of $d$ that will be set in
the analysis, we construct a $2t\times n$ matrix $M$ in a column-wise fashion.
For each column $j$ of $M$, we choose $t/d$
rows at random and we set the values of these entries to $1$.
The other entries in
column $j$ are set to $0$.
In other words, we ``inject'' the sample $j$ into each of the $t/d$
random tests we pick for the corresponding column (since
rows of $M$ correspond to tests and the columns correspond to
samples).
Note, then, that for any set of $d$ defective samples,
there are at most $t$ tests that will have positive outcomes and,
therefore, at least $t$ tests that will have negative outcomes.
The columns that correspond to samples that are distinguishable from
the defectives ones can be immediately identified.
The remaining issue, then, is to determine the value of $t$ needed so
that, for a given value of $k$, $M$ is a $(d,k)$-resolvable matrix
with high probability.

Let $D$ be a fixed set of $d$ defectives samples.
For each (column) item $i$ in $C-D$, let $X_i$ denote the
indicator random variable that is $1$ if $i$ is falsely identified as
a positive sample by $M$ (that is, $i$ is not included in the set of
(negative) items distinguished from those in $D$), and is $0$
otherwise.
Observe that the $X_i$'s are independent, since $X_i$ depends only on
whether the choice of rows we picked for column $i$ collide with the
at most $t$ rows of $M$ that we picked for the columns corresponding
to items in $D$.
Furthermore, this observation implies that any $X_i$ is $1$
(a false positive)
with probability at most $2^{-t/d}$.
Therefore, the expected value of $X$, $E[X]$, is at most
${\hat\mu}=n/2^{t/d}$.
This fact allows us to apply Lemma~\ref{lem:chernoff} to bound the
probability that $M$ does not satisfy the $(d,k)$-resolvable property
for this particular choice, $D$, of $d$ defective samples.
In particular,
\[
\Pr(X\ge k) \le \left(\frac{e{\hat\mu}}{k}\right)^k =
\frac{\left(\frac{en}{k}\right)^k}{2^{(t/d)k}}
.
\]
Note that this bound immediately implies that if $k=1$ and $t\ge d(e+1)\log n$,
then $M$ will be completely $(d,1)$-resolvable with high probability
($1-1/n$) for any particular set of defective items, $D$.

We are interested, however, in a bound 
implying that for \emph{any} subset $D$ of $d$ defectives (of which
there are ${n\choose d}<(en/d)^d$, by Lemma~\ref{lem:choose}),
our matrix $M$ is $(d,k)$-resolvable with high probability,
that is, probability at least $1-1/n$.
That is,
we are interested in the value of $t$ such that the above probability
bound is $(en/d)^{-d}/n$.  From the above 
probability bound, therefore, we are interested in a
value of $t$ such that
\ifFull
\[
\frac{2^{(t/d)k}}{\left(\frac{en}{k}\right)^k}  \ge
\left(\frac{en}{d}\right)^d n .
\]
That is, we would like
\fi
\[
2^{(t/d)k} \ge
\left(\frac{en}{d}\right)^d
\left(\frac{en}{k}\right)^k
n .
\]
This bound will hold whenever
\ifFull
\[
t \ge (d^2/k) \log (en/d) + d\log (en/k) + (d/k)\log n .
\]
\else
$t \ge (d^2/k) \log (en/d) + d\log (en/k) + (d/k)\log n $.
\fi
Thus, we have the following.

\begin{theorem}
If $ t \ge (d^2/k) \log (en/d) + d\log (en/k) + (d/k)\log n $,
then a $2t\times n$ random matrix $M$
constructed by sample-injection
is $(d,k)$-resolvable with high probability, that is, with
probability at least $1-1/n$.
\end{theorem}

\ifFull
Taking $k=1$, therefore, we have an alternative method for
constructing a $d$-disjunct matrix $M$ with high probability:

\begin{corollary}
If $ t \ge d^2 \log (en/d) + d\log en + d\log n $,
then a $2t\times n$ random matrix $M$ constructed by
sample-injection is $d$-disjunct with high probability.
\end{corollary}

That is, we can construct a one-round group test based on
sample-injection that uses $O(d^2\log (n/d))$ tests.
\fi

As mentioned above, a productive
way of using the sample-injection construction is
to build a $(d,k)$-resolvable matrix $M$ for a reasonably small value
of $k$.
We can then use this matrix as the first round in a two-round 
rake-and-winnow testing
strategy, where the second round simply
involves our individual testing of the at most $d+k$ samples left as
potential positive samples from the first round.

\begin{corollary}
If $ t \ge 2d \log (en/d) + \log n $,
then the $2t\times n$ random matrix $M$ constructed by
sample-injection is $(d,d)$-resolvable with high probability.
\end{corollary}

This corollary implies that we can construct a rake-and-winnow
algorithm where the first stage involves performing $O(d\log (n/d))$
tests, which is within a (small) constant factor of the information
theoretic lower bound, and the second round involves individually testing
at most $2d$ samples.

\section{Improved Bounds for Small $d$ Values}
In this section, we consider efficient algorithms for
the special cases when $d=2$ and $d=3$. 
We present time-optimal algorithms for these cases;
that is, with $A(n,t)$ being $O(t)$.
Our algorithm for $d=3$ is the first known such algorithm.

\paragraph{Finding up to Two Defectives.}
Consider the problem of determining which items are defective
when we know that there are at most two defectives.
We describe a $\overline 2$-separable matrix and a
time-optimal analysis algorithm with $t = (q^2 +5q)/2$
and $n = 3^q$, for any positive integer $q$.

Let the number of items be $n = 3^q$,
and let the item indices be expressed in radix 3.
Index $X = X_{q-1} \cdots X_0$, where each digit $X_p \in \{0,1,2\}$.

Hereafter,
$X$ ranges over the item index numbers $\{0, \ldots n-1\}$,
$p$ ranges over the radix positions $\{0, \ldots q-1\}$,
and $v$ ranges over the digit values $\{0,1,2\}$.

For our construction, matrix $M$ is partitioned into submatrices $B$ and $C$.
Matrix $B$ is the submatrix of $M$ consisting of its first $3q$ rows.
Row $\langle p,v\rangle$ of $B$ is associated with radix position $p$ and value $v$.
$B[\langle p,v\rangle,X] = 1$ iff $X_p = v$.

Matrix $C$ is the submatrix of $M$ consisting of its last $q \choose 2$ rows.
Row $\langle p, p'\rangle$ of $C$ is associated with distinct radix positions $p$ and $p'$,
where $p < p'$.  $C[\langle p, p'\rangle,X] = 1$ iff $X_p = X_{p'}$.

Let $test_B(p,v)$ be the result (1 for positive, 0 for negative)
of the test of items having a 1-entry in row $\langle p,v\rangle$ in $B$.
Similarly, let $test_C(p, p')$ be the result of testing row $\langle p, p'\rangle$ in $C$.
Let $test1(p)$ be the number of different values held
by defectives in radix position $p$.
$test1(p)$ can be computed by $test_B(p,0) + test_B(p,1) + test_B(p,2)$.

The analysis algorithm is shown
in the Appendix
in Figure~\ref{fig:Analysis}.

It is easy to determine how many defective items are present.
There are no defective items when $test1(0)=0$.
There is only one defective item when $test1(p)=1$ for all $p$,
since if there were two defective items then there must be
at least one position $p$ in which their indices differ
and $test1(p)$ would then have value 2.
The one defective item has index $D = D_{q-1} \cdots D_0$,
where digit $D_p$ is the value $v$ for which $test_B(p,v) = 1$.

Otherwise, there must be 2 defective items,
$D = D_{q-1} \cdots D_0$ and $E = E_{q-1} \cdots E_0$.
We iteratively determine the values of the digits of indices $D$ and $E$.

For radix positions in which defective items exist for only one value
of that digit, both $D$ and $E$ must have that value for that digit.
For each other radix position, two distinct values
for that digit occur in the defective items.

The first radix position in which $D$ and $E$
differ is recorded in the variable $p^{*}$
and the value of that digit in $D$ (respectively, $E$) is recorded
in $v^{*}_1$ (respectively, $v^{*}_2$).

For any subsequent position $p$ in which $D$ and $E$ differ,
the digit values of the defectives in that position are
$v_a$ and $v_b$, which are two distinct values from $\{0,1,2\}$,
as are $v^{*}_1$ and $v^{*}_2$,
and therefore there must be at least one value in common between
$\{v_a,v_b\}$ and $\{v^{*}_1,v^{*}_2\}$.

Let a common value be $v_a$ and, without loss of generality,
let $v_a = v^{*}_1$.

\begin{lemma}
\ifFull
The digit assignment for position $p$
is $D_p=v_a$ and $E_p=v_b$ iff $test_C(p^{*},p)=1$.
\else
The digit assignment for $p$
is $D_p=v_a$ and $E_p=v_b$ iff $test_C(p^{*},p)=1$.
\fi
\end{lemma}

\ifFull
\begin{proof}
We consider the two possibilities of which defective item has $v_a$
as its digit in position $p$.

{\it Case 1.} $D_p=v_a$.\\
We see that $D_p=v_a=v^{*}_1$.
Accordingly, a defective ($D$) would be among the
items tested in $test_C(p^{*},p)$.
Therefore, $test_C(p^{*},p) = 1$.

{\it Case 2.} $E_p=v_a$.\\
We see that $D_p \neq v^{*}_1$,
because $D_p \neq E_p = v_a = v^{*}_1$,
and also that $E_p \neq v^{*}_2$,
because $E_p = v_a = v^{*}_1 \neq v^{*}_2$.
Accordingly, neither of the defective items would be among the items
tested in $test_C(p^{*},p)$.
Therefore, $test_C(p^{*},p) = 0$.
\end{proof}
\fi

We have determined the values of defectives D and E for all positions --
those where they are the same and those where they differ.
For each position, only a constant amount of work is required to
determine the assignment of digit values.
Therefore, we have proven the following theorem.

\begin{theorem}
A $\overline 2$-separable matrix
that has a time-optimal analysis algorithm
can be constructed with $t = (q^2 +5q)/2$
and $n = 3^q$, for any positive integer $q$.
\end{theorem}

\paragraph{Comparison of the Number of Tests Required for $d=2$ Method.}
\ifFull
A $\overline 2$-separable or a 2-disjunct
$t \times n$ matrix enables determination of up to 2
defective items from among $n$ {\em or fewer} items using $t$ tests.
An algorithm is more competitive at or just below one of its breakpoints,
values of $n$ for which increasing $n$ by one significantly increases $t$.
The MR algorithm has breakpoints at one under all powers of 2,
our ($d$=2) algorithm at all powers of 3,
and the KS algorithm at only certain powers of 3.
Our general-$d$ algorithms do not have significant breakpoints.

Table~\ref{tbl:Various} lists the number of tests required by these
algorithms for some small values of $n$.
\fi
For all $n \leq 3^{63}$, our $d=2$ algorithm uses the smallest number of tests.
For higher values of $n \leq 3^{130}$,
the Kautz/Singleton and our $d=2$ and general (Chinese Remainder Sieve)
algorithms alternate being dominant.
\ifFull
The alternations are illustrated in Table~\ref{tbl:Dominant}.
\fi
For all $n \geq 3^{131}$,
the Hwang/S\'{o}s algorithm uses the fewest tests.

\ifFull
\begin{table}[htp]
\caption[]{$t(n)$ for small $n$ ($d=2$)}
\[
\begin{array}{|l|rrrrrrrrrr|}
\hline
(d=2)  & 15 &100 &10^3&10^4 &10^5 &10^6 &10^8 &10^{10}&10^{20}&10^{30}\\
\hline
{\rm our\ } d=2
       & 12 & 25 & 42 &  63 &  88 & 117 & 187 & 273   &  987 & 2142 \\
{\rm our\ bktrk}
       & 19 & 36 & 60 &  89 & 131 & 168 & 268 & 378   & 1176 & 2350 \\
{\rm our\ genl}
       & 28 & 41 & 77 & 100 & 160 & 197 & 281 & 440   & 1264 & 2584 \\
{\rm MR}&14 & 35 & 65 & 119 & 170 & 230 & 405 & 629   & 2345 & 5150 \\
{\rm KS}&27 & 81 & 81 & 243 & 243 & 243 & 729 & 729   & 2187 & 2187 \\
{\rm HS}&   &373 &507 & 641 & 775 & 909 &1177 &1446   & 2787 & 4129 \\
\hline
\end{array}
\]
\label{tbl:Various}
\end{table}

\begin{table}[htp]
\caption[]{$t(n)$ for large $n$ ($d=2$)}
\[
\begin{array}{|l|rrrrrrr|}
\hline
(d=2)            &    3^{63} &   3^{64} &   3^{104} &   3^{112}
                       &  3^{128} &  3^{130} &  3^{256} \\
\hline
{\rm our\ }d=2   & {\bf 2142}&     2208 & {\bf 5668}&     6552
                       &     8512 &     8775 &    33408 \\
{\rm our\ bktrk} &      2366 &     2424 &      5687 &{\bf 6454}
                       &     8184 &{\bf 8394}&    28311 \\
{\rm our\ genl}  &      2584 &     2584 &      6081 &     6870
                       &     8582 &     8893 &    29296 \\
{\rm KS}         &      2187 &{\bf 2187}&      6561 &     6561
                       &{\bf 6561}&    19683 &    19683 \\
{\rm HS}         &      4136 &     4200 &      6760 &     7272
                       &     8296 &     8424 &{\bf16488}\\
\hline
\end{array}
\]
\label{tbl:Dominant}
\end{table}
\fi

\paragraph{Finding up to Three Defectives.}
Consider the problem of determining which items are defective
when we know that there are at most three defectives.
We describe a $\overline 3$-separable matrix
and a time-optimal analysis algorithm
with $t = 2q^2 -2q$ and $n = 2^q$, for any positive integer $q$.

Let the number of items be $n = 2^q$,
and let the item indices be expressed in radix 2.
Index $X = X_{q-1} \cdots X_0$, where each digit $X_p \in \{0,1\}$.

Hereafter,
$X$ ranges over the item index numbers $\{0, \ldots n-1\}$,
$p$ ranges over the radix positions $\{0, \ldots q-1\}$,
and $v$ ranges over the digit values $\{0,1\}$.

Matrix $M$ has $2q^2 -2q$ rows.
Row $\langle p, p', v, v'\rangle$ of $M$ is associated with
distinct radix positions $p$ and $p'$, where $p < p'$,
and with values $v$ and $v'$, each of which is in \{0,1\}.
$M[\langle p, p', v, v'\rangle, X] = 1$ iff
$X_p = v$ and $X_{p'} = v'$.

Let $test_M(p, p', v, v')$ be the result
(1 for positive, 0 for negative) of testing items having
a 1-entry in row $\langle p, p', v, v'\rangle$ in $M$.
For $p' > p$, define $test_M(p', p, v', v)$ =
$test_M(p, p', v, v')$.

The following three functions can be computed in terms of $test_M$.
\begin{itemize}
\item $test_B(p,v)$ has value 1 (0) if there are (not)
any defectives having value $v$ in radix position $p$.
Hence,
$test_B(0,v) = 0$ if $test_M(0,1,v,0) + test_M(0,1,v,1) = 0$,
and 1 otherwise.
For $p>0$, $test_B(p,v) = 0$ if $test_M(p,0,v,0) + test_M(p,0,v,1) = 0$,
and 1 otherwise.

\item $test1(p)$ is the number of different binary values held
by defectives in radix position $p$.
Thus,
$test1(p) = test_B(p,0) + test_B(p,1)$.

\item $test2(p, p')$ is
the number of different ordered pairs of binary values held
by defectives in the designated ordered pair of radix positions.
\ifFull

$test2(p, p') = test_M(p, p', 0, 0) + test_M(p, p', 0, 1) +
test_M(p, p', 1, 0) + test_M(p, p', 1, 1)$.
\else
Therefore,
{
$test2(p, p') = test_M(p, p', 0, 0) + test_M(p, p', 0, 1) +
test_M(p, p', 1, 0) + test_M(p, p', 1, 1)$.
}
\fi
\end{itemize}

The analysis algorithm is shown
in the Appendix
in Figure~\ref{fig:Analysis3}.

We determine the number of defective items and the value of their digits.
There are no defective items when $test1(0)=0$.
At each radix position $p$ in which $test1(p)=1$,
all defective items have the same value of that digit.
If all defectives agree on all digit values, then there is only
one defective.
Otherwise there are at least two defectives, and we need to consider
how to assign digit values for only the set of positions $P$
in which there is at least one defective
having each of the two possible binary digit values.

\begin{lemma}
There are only two defectives if and only if,
for $p,p' \in P, test2(p,p') = 2$.
\end{lemma}

\ifFull
\begin{proof}
A defective item can contribute at most one new combination of
values in positions $p,p'$ and so $test2(p,p') \leq$
the number of defectives.
Accordingly, if there are fewer than two defectives then
$test2(p,p') < 2$.

If there are exactly two defectives then
$test2(p,p') \leq 2$.
Since $p \in P$, both binary values appear among defectives,
so $test2(p,p') \geq 2$, and therefore $test2(p,p') = 2$. 

Consider the case in which there are three defectives.
In any position $p_1$ in which both binary values appear
at that digit among the set of defectives, one of the defectives
(say, $D$)
has one binary value (say, $v_1$)
and the other two defectives
($E, F$)
have the other binary value ($\overline v_1$).
Since $E$ and $F$ are distinct,
they must differ in value at some other position $p_2$.
Therefore, there will be three different ordered pairs of binary values
held by defectives in positions $p_1$ and $p_2$,
and so $test2(p_1,p_2) = 3$.
\end{proof}
\fi

Accordingly, if there is no pair of positions for which $test2$
has value 3, we can conclude that there are only two defectives.
Otherwise, there are positions $p_1,p_2$ for which $test2(p_1, p_2) = 3$,
and one of the four combinations of two binary values will not appear.
Let that missing combination be $v_1,v_2$.
Thus, while position $p_1$ uniquely identifies one defective, say $D$,
as the only defective having value $v_1$ at that position,
position $p_2$ uniquely identifies one of the other
defectives, say $E$, as having value $v_2$.

\begin{lemma}
If the position $p^*$ uniquely identifies
the defective $X$ to have value $v^*$,
then the value of the defective $X$ at any other position $p$
will be that value $v$ such that $test_M(p^*, p, v^*, v)=1$.
\end{lemma}

\ifFull
\begin{proof}
If position $p^*$ uniquely identifies
defective $X$ as having value $v^*$,
then $X_{p^*} = v^*$ and, for any other defective $Y$,
$Y_{p^*} \neq v^*$.

Let $v = X_p$, for any $p\neq p^*$.
Then $test_M(p^*, p, v^*, v)=1$, since $X$ is a defective
that has the required values at the required positions
to be included in this test.

Also, $test_M(p^*, p, v^*, \overline v)=0$,
because none of the defectives are included in this test.
Defective $X$ is not included because $X_p \neq \overline v$.
Any other defective, $Y \neq X$, is not included because
$Y_{p^*} \neq v^*$.
\end{proof}
\fi

Since we have positions that uniquely identify $D$ and $E$,
we can determine the values of all their other digits and
the only remaining problem is to determine the values of the digits
of defective $F$.

Since position $p_1$ uniquely identifies $D$,
we know that $F_{p_1} = \overline v_1$.
For any other position $p$,
after determining that $E_p = v$,
we note that if
$test_M(p_1,p,\overline v_1,\overline v) = 1$
then there must be at least one defective, $X$,  for which
$X_{p_1}=\overline v_1$ and $X_p=\overline v$.
Defective $D$ is ruled out since $D_{p_1} = v_1$, and
defective $E$ is ruled out since $E_p = v$.
Therefore, it must be that $F_p = \overline v$.
Otherwise, if that $test_M = 0$ then $F_p = v$,
since $F_p = \overline v$ would have caused $test_M = 1$.

We have determined the values of defectives D, E and F for all positions.
For each position, only a constant amount of work is required to
determine the assignment of digit values.
Therefore, we have proven the following theorem.

\begin{theorem}
A $\overline 3$-separable matrix
that has a time-optimal analysis algorithm
can be constructed with $t = 2q^2 -2q$
and $n = 2^q$, for any positive integer $q$.
\end{theorem}

\paragraph{Comparison of the Number of Tests Required for $d=3$ Method.}
The general $d$ algorithm due to Hwang and S\'{o}s \cite{HS87}
requires fewer tests than does the algorithm
for $d=3$ suggested by Du and Hwang \cite{dh-cgtia-00}.
For $n<10^{10}$,
our ($d=3$) algorithm requires even fewer tests
and our general (Chinese Remainder Sieve) algorithm fewest.
However, asymptotically Hwang/S\'{o}s uses the fewest tests.
We note that, unlike these other efficient algorithms,
our ($d=3$) algorithm is time-optimal.
\ifFull
Table~\ref{tbl:Compare3} lists the number of tests required by
these algorithms for some small values of $n$.

\begin{table}[htp]
\caption[]{Comparing $t(n)$ for $d=3$}
\[
\begin{array}{|l|rrrrrrr|}
\hline
(d=3)         & 100 & 10^4& 10^6 & 10^8 & 10^{10}& 10^{20}& 10^{30}\\
\hline
{\rm our\ bktrk}
              &  60 & 168 &  321 &  513 &   738  &  2350  &   4777 \\
{\rm our\ genl}
              &  77 & 197 &  381 &  568 &   791  &  2584  &   5117 \\
{\rm our\ }d=3&  84 & 364 &  760 & 1404 &  2244  &  8844  &  19800 \\
{\rm HS}      & 838 &1442 & 2046 & 2649 &  3253  &  6271  &   9289 \\
{\rm DH}      & 840 &3444 & 7080 &12960 & 20604  & 80400  & 179400 \\
\hline
\end{array}
\]
\label{tbl:Compare3}
\end{table}
\fi

\paragraph{\textbf{Acknowledgments.}}
We would like to thank
George Lueker
and
Dennis Shasha
for several helpful discussions related to the topics of this paper.
This work was supported in part by NSF Grants CCR-0312760,
CCR-0311720, CCR-0225642, and CCR-0098068.
The results of this paper were 
announced in preliminary form in~\cite{egh-icgtr-05}.

\bibliographystyle{abbrv}
\bibliography{cgt,goodrich}

\ifFull
\newpage
\appendix
\section{Pseudo-Code Listings}
\fi

\ifFull
\begin{figure}[hb!]
\rule{4.0in}{0.01in}
\begin{lstlisting}
def eratosthenes():
    """Generate the sequence of prime numbers via the Sieve of Eratosthenes."""
    D = {}  # map composite integers to primes witnessing their compositeness
    q = 2   # first integer to test for primality
    while True:
        if q not in D:
            yield q        # not marked composite, must be prime
            D[q*q] = [q]   # first multiple of q not already marked
        else:
            for p in D[q]: # move each witness to its next multiple
                D.setdefault(p+q,[]).append(p)
            del D[q]       # no longer need D[q], free memory
        q += 1

def search(primes,maxpow,target):
    """
    Backtracking search for exponents of prime powers, each at most maxpow,
    so that the product of the powers is at least target and the sum of the
    non-unit powers is minimized.  Returns the pair [sum,list of exponents].
    """
    if target <= 1:         # all unit powers will work?
        return [0,[0]*len(primes)]
    elif not primes or maxpow**len(primes) < target:
        return None         # no primes supplied, no solution exists
    primes = list(primes)   # list all but the last prime for recursive calls
    p = primes.pop()
    best = None             # no solution found yet
    i = 0
    while p**i <= maxpow:   # loop through possible exponents of p
        s = search(primes,maxpow,(target + p**i - 1)//p**i)
        if s is not None:
            s[0] += i and p**i
            s[1].append(i)
            best = min(best,s) or s
        i += 1
    return best
\end{lstlisting}
\rule{4.0in}{0.01in}
\caption{Subroutines for construction based on prime factorization}
\label{fig:Primes-subs}
\end{figure}

\begin{figure}[htp]
\rule{4.0in}{0.01in}
\begin{lstlisting}
def prime_cgt(n,d):
    """Find a CGT for n and d and output a description of it to stdout."""

    # collect primes until their total product is large enough
    primes = []
    product = 1
    for p in eratosthenes():
        primes.append(p)
        product *= p
        if product > n**d:
            break
    
    # now find good collection of powers of those primes...
    result = search(primes,primes[-1],n**d)
    powers = result[1]

    # output results
    print "n =",n,"d =",d,":",
    for i in range(len(primes)):
        if powers[i] == 1:
            print primes[i],
        elif powers[i] > 1:
            print str(primes[i]) + "^" + str(powers[i]),
    print "total tests:", sum([primes[i]**powers[i] for i in range(len(primes))
                                if powers[i]])

if __name__ == "__main__":
    for d in range(2,6):
        for x in range(6,16):
            prime_cgt(1<<x,d)
        print
\end{lstlisting}
\rule{4.0in}{0.01in}
\caption{Construct tests based on prime factorization}
\label{fig:Primes}
\end{figure}
\fi

\ifFull
\begin{figure}[ht]
\rule{4.0in}{0.01in}
\begin{tabbing}
aaaa \= aaaa \= aaaa \= aaaa \= aaaa \= aaaa \= aaaa \= \kill

\> {\bf if} $test1(0)=0$ {\bf then return} there are no defective items \\
\> $p^{*} \leftarrow -1$ \\
\> {\bf for} $p \leftarrow 0$ {\bf to} $q-1$ {\bf do} \\
\> \> {\bf if} $test1(p)=1$ {\bf then} \\
\> \> \> $D_p \leftarrow E_p \leftarrow$ the value $v$ such that $test_B(p,v)=1$ \\
\> \> {\bf else} \> \> // $test1(p)$ has value 2 \\
\> \> \> Let $v_1,v_2$ be the two values of $v$ such that $test_B(p,v)=1$ \\
\> \> \> {\bf if} $p^{*} < 0$ {\bf then} \\
\> \> \> \>  $p^{*} \leftarrow p$ \\
\> \> \> \>  $v^{*}_1 \leftarrow D_p \leftarrow v_1$ \\
\> \> \> \>  $v^{*}_2 \leftarrow E_p \leftarrow v_2$ \\
\> \> \> {\bf else} \\
\> \> \> \>  {\bf if} $test_C(p^{*},p)=1$ {\bf and} ( $v^{*}_1=v_1$ {\bf or} $v^{*}_2=v_2$ ) {\bf then} \\
\> \> \> \> \>  $D_p \leftarrow v_1$ \\
\> \> \> \> \>  $E_p \leftarrow v_2$ \\
\> \> \> \>  {\bf else} \\
\> \> \> \> \>  $D_p \leftarrow v_2$ \\
\> \> \> \> \>  $E_p \leftarrow v_1$ \\
\> {\bf if} $p^{*} < 0$ {\bf then} \\
\> \> {\bf return} there is one defective item $D$ \\
\> {\bf else} \\
\> \> {\bf return} there are two defective items $D$ and $E$
\end{tabbing}
\rule{4.0in}{0.01in}
\caption{Analysis algorithm for up to 2 defectives}
\label{fig:Analysis}
\end{figure}

\begin{figure}[htp]
\rule{4.0in}{0.01in}
\begin{tabbing}
aaaa \= aaaa \= aaaa \= aaaa \= aaaa \= aaaa \= aaaa \= \kill

\> {\bf if} $test1(0)=0$ {\bf then return} there are no defective items \\
\> $P \leftarrow \emptyset$ \\
\> {\bf for} $p \leftarrow 0$ {\bf to} $q-1$ {\bf do} \\
\> \> {\bf if} $test1(p)=1$ {\bf then} \\
\> \> \> $D_p \leftarrow E_p \leftarrow F_p \leftarrow$ the value $v$ s.t. $test_B(p,v)=1$ \\
\> \> {\bf else} \> $P \leftarrow P \cup \{p\}$ \\
\> {\bf if} $P=\emptyset$ {\bf then return} there is one defective item $D$ \\
\> {\bf if} $test2(p_1, p_2)=2$ for all $p_1, p_2 \in P$ {\bf then} \\
\> \> $p^{*} \leftarrow -1$ \\
\> \> {\bf for} $p \in P$ {\bf do} \\
\> \> \> {\bf if} $p^{*} < 0$ {\bf then} \\
\> \> \> \>  $p^{*} \leftarrow p$ \\
\> \> \> \>  $v^{*} \leftarrow D_p \leftarrow 0$ \\
\> \> \> {\bf else if} $test_M(p^{*},p,v^{*},0)=1$ {\bf then} \\
\> \> \> \> \>  $D_p \leftarrow 0$ \\
\> \> \> \>  {\bf else} \> $D_p \leftarrow 1$ \\
\> \> \>  $E_p \leftarrow 1-D_p$ \\
\> \> {\bf return} there are two defective items $D, E$ \\
\> {\bf else} \\
\> \> Let $p_1, p_2$ be positions such that $test2(p_1, p_2)=3$ \\
\> \> Let $v_1, v_2$ be values such that $test_M(p_1, p_2, v_1, v_2)=0$ \\
\> \> $D_{p_1} \leftarrow v_1$ \\
\> \> $F_{p_1} \leftarrow E_{p_1} \leftarrow 1-v_1$ \\
\> \> $E_{p_2} \leftarrow v_2$ \\
\> \> $F_{p_2} \leftarrow D_{p_2} \leftarrow 1-v_2$ \\
\> \> {\bf for} $p \in P-\{p_1,p_2\}$ {\bf do} \\
\> \> \> {\bf if} $test_M(p_1, p, v_1, 0)=1$ {\bf then} \\
\> \> \> \> $D_p \leftarrow 0$ \\
\> \> \> {\bf else} \> $D_p \leftarrow 1$ \\
\> \> \> {\bf if} $test_M(p_2, p, v_2, 0)=1$ {\bf then} \\
\> \> \> \> $E_p \leftarrow 0$ \\
\> \> \> {\bf else} \> $E_p \leftarrow 1$ \\
\> \> \> $v \leftarrow E_p$ \\
\> \> \> {\bf if} $test_M(p_1, p, 1-v_1, 1-v)=1$ {\bf then} \\
\> \> \> \> $F_p \leftarrow 1-v$ \\
\> \> \> {\bf else} \> $F_p \leftarrow v$ \\
\> {\bf return} there are three defective items $D, E$, and $F$
\end{tabbing}
\rule{4.0in}{0.01in}
\caption{Analysis algorithm for up to 3 defectives}
\label{fig:Analysis3}
\end{figure}

\else   

\begin{figure}[htp]
\begin{footnotesize}
\begin{tabular}{l|l}
\begin{minipage}{2.3in}
\begin{tabbing}
\= a \= a \= a \= a \= a \= a \= \kill

\> {\bf if} $test1(0)=0$ {\bf then} \\
\>\> {\bf return} there are no defective items \\
\> $p^{*} \leftarrow -1$ \\
\> {\bf for} $p \leftarrow 0$ {\bf to} $q-1$ {\bf do} \\
\> \> {\bf if} $test1(p)=1$ {\bf then} \\
\> \> \> Let $D_p$ and $E_p$ be the (same) \\
\> \> \> \> value $v$ such that $test_B(p,v)=1$ \\
\> \> {\bf else} // $test1(p)$ has value 2 \\
\> \> \> Let $v_1,v_2$ be the two values  \\
\> \> \> \> of $v$ such that $test_B(p,v)=1$ \\
\> \> \> {\bf if} $p^{*} < 0$ {\bf then} \\
\> \> \> \>  $p^{*} \leftarrow p$ \\
\> \> \> \>  $v^{*}_1 \leftarrow D_p \leftarrow v_1$ \\
\> \> \> \>  $v^{*}_2 \leftarrow E_p \leftarrow v_2$ \\
\> \> \> {\bf else} \\
\> \> \> \>  {\bf if} $test_C(p^{*},p)=1$ \\
\> \> \> \> \> {\bf and} ( $v^{*}_1=v_1$ {\bf or} $v^{*}_2=v_2$ ) {\bf then} \\
\> \> \> \> \>  $D_p \leftarrow v_1$ \\
\> \> \> \> \>  $E_p \leftarrow v_2$ \\
\> \> \> \>  {\bf else} \\
\> \> \> \> \>  $D_p \leftarrow v_2$ \\
\> \> \> \> \>  $E_p \leftarrow v_1$ \\
\> {\bf if} $p^{*} < 0$ {\bf then} \\
\> \> {\bf return} one defective, $D$ \\
\> {\bf else} \\
\> \> {\bf return} two defectives, $D$ and $E$
\end{tabbing}
\end{minipage}
&
\begin{minipage}{2.3in}
\begin{tabbing}
a\= a \= a \= a \= a \= a \= a \= \kill

\> {\bf if} $test1(0)=0$ {\bf then} \\
\> \> {\bf return} there are no defective items \\
\> $P \leftarrow \emptyset$ \\
\> {\bf for} $p \leftarrow 0$ {\bf to} $q-1$ {\bf do} \\
\> \> {\bf if} $test1(p)=1$ {\bf then} \\
\> \> \> Let $D_p$,  $E_p$, and $F_p$ be the (same) \\
\> \> \> \> value $v$ such that $test_B(p,v)=1$ \\
\> \> {\bf else} $P \leftarrow P \cup \{p\}$ \\
\> {\bf if} $P=\emptyset$ {\bf then return} there is one defective item $D$ \\
\> {\bf if} $test2(p_1, p_2)=2$ for all $p_1, p_2 \in P$ {\bf then} \\
\> \> $p^{*} \leftarrow -1$ \\
\> \> {\bf for} $p \in P$ {\bf do} \\
\> \> \> {\bf if} $p^{*} < 0$ {\bf then} \\
\> \> \> \>  $p^{*} \leftarrow p$ \\
\> \> \> \>  $v^{*} \leftarrow D_p \leftarrow 0$ \\
\> \> \> {\bf else if} $test_M(p^{*},p,v^{*},0)=1$ {\bf then} \\
\> \> \> \> \>  $D_p \leftarrow 0$ \\
\> \> \> \>  {\bf else} $D_p \leftarrow 1$ \\
\> \> \>  $E_p \leftarrow 1-D_p$ \\
\> \> {\bf return} there are two defective items $D, E$ \\
\> {\bf else} \\
\> \> Let $p_1, p_2$ be positions s.t. $test2(p_1, p_2)=3$ \\
\> \> Let $v_1, v_2$ be values s.t. $test_M(p_1, p_2, v_1, v_2)=0$ \\
\> \> $D_{p_1} \leftarrow v_1$ \\
\> \> $F_{p_1} \leftarrow E_{p_1} \leftarrow 1-v_1$ \\
\> \> $E_{p_2} \leftarrow v_2$ \\
\> \> $F_{p_2} \leftarrow D_{p_2} \leftarrow 1-v_2$ \\
\> \> {\bf for} $p \in P-\{p_1,p_2\}$ {\bf do} \\
\> \> \> {\bf if} $test_M(p_1, p, v_1, 0)=1$ {\bf then} \\
\> \> \> \> $D_p \leftarrow 0$ \\
\> \> \> {\bf else} $D_p \leftarrow 1$ \\
\> \> \> {\bf if} $test_M(p_2, p, v_2, 0)=1$ {\bf then} \\
\> \> \> \> $E_p \leftarrow 0$ \\
\> \> \> {\bf else} $E_p \leftarrow 1$ \\
\> \> \> $v \leftarrow E_p$ \\
\> \> \> {\bf if} $test_M(p_1, p, 1-v_1, 1-v)=1$ {\bf then} \\
\> \> \> \> $F_p \leftarrow 1-v$ \\
\> \> \> {\bf else} $F_p \leftarrow v$ \\
\> {\bf return} there are three defective items $D, E$, and $F$
\end{tabbing}
\end{minipage}
\\[10pt]
\hspace*{1in}(a) & \hspace*{1in}(b)
\end{tabular}
\end{footnotesize}
\caption{Analysis algorithms. (a) for up to 2 defectives; (b)
 for up to 3 defectives.}
\label{fig:Analysis}
\label{fig:Analysis3}
\end{figure}
\fi

\end{document}